\documentclass[aps,pra,twocolumn,superscriptaddress]{revtex4}
\usepackage{amsmath}
\usepackage{graphicx}
\usepackage[section]{placeins}
\usepackage{MnSymbol}
\usepackage{enumitem}
\usepackage{color}
\usepackage{times}%
\newcommand{\op}[1]{\widehat{#1}}
\newcommand{\ket}[1]{|#1\rangle}
\newcommand{\bra}[1]{\langle#1|}

\begin{document}
\title{Hardy's paradox tested in the spin-orbit Hilbert space of single photons}
\author{Ebrahim Karimi}
\email{ekarimi@uottawa.ca}
\affiliation{Dipartimento di Fisica, Universit\`{a} di Napoli Federico II, Compl.\ Univ.\ di Monte S. Angelo, 80126 Napoli, Italy}
\affiliation{Department of Physics, University of Ottawa, 150 Louis Pasteur, Ottawa, Ontario, K1N 6N5 Canada}
\author{Filippo Cardano}
\affiliation{Dipartimento di Fisica, Universit\`{a} di Napoli Federico II, Compl.\ Univ.\ di Monte S. Angelo, 80126 Napoli, Italy}
%
\author{Maria Maffei}
\affiliation{Dipartimento di Fisica, Universit\`{a} di Napoli Federico II, Compl.\ Univ.\ di Monte S. Angelo, 80126 Napoli, Italy}
%
\author{Corrado de Lisio}
\affiliation{Dipartimento di Fisica, Universit\`{a} di Napoli Federico II, Compl.\ Univ.\ di Monte S. Angelo, 80126 Napoli, Italy}
%
\author{Lorenzo Marrucci}
\affiliation{Dipartimento di Fisica, Universit\`{a} di Napoli Federico II, Compl.\ Univ.\ di Monte S. Angelo, 80126 Napoli, Italy}
\affiliation{CNR-SPIN, Compl.\ Univ.\ di Monte S. Angelo, 80126 Napoli, Italy}
\author{Robert W. Boyd}
\affiliation{Department of Physics, University of Ottawa, 150 Louis Pasteur, Ottawa, Ontario, K1N 6N5 Canada}
%
\author{Enrico Santamato}
\email{santamato@na.infn.it}
\affiliation{Dipartimento di Fisica, Universit\`{a} di Napoli Federico II, Compl.\ Univ.\ di Monte S. Angelo, 80126 Napoli, Italy}
\affiliation{Consorzio Nazionale Interuniversitario per le Scienze Fisiche della Materia, Napoli}
\begin{abstract}
We test experimentally the quantum ``paradox'' proposed by Lucien Hardy in 1993 by using single photons instead of photon pairs. This is achieved by addressing two compatible degrees of freedom of the same particle, namely its spin angular momentum, determined by the photon polarization, and its orbital angular momentum, a property related to the optical transverse mode. Because our experiment involves a single particle, we cannot use locality to logically enforce non-contextuality, which must therefore be assumed based only on the observables' compatibility. On the other hand, our single-particle experiment can be implemented more simply and allows larger detection efficiencies than typical two-particle ones, with a potential future advantage in terms of closing the detection loopholes.
\end{abstract}
\maketitle
\section{Introduction}
Since the 1935 famous publication by Einstein, Podolsky, and Rosen (EPR) \cite{EPR35}, which posed the problem in its clearest form, the quantum theory has repeatedly eluded any simple realistic interpretation. Although EPR actually aimed at demonstrating quantum theory incompleteness, the subsequent celebrated Bell's theorem showed that no possible theory assuming additional ``hidden variables'' in order to describe the underlying reality could be in agreement with quantum predictions, provided one also made the very plausible assumption of ``locality'' \cite{Be64} for the behavior of such hidden variables. The difference between the predictions of quantum theory and those of all possible local realistic theories could be quantified in well defined ``inequalities'' concerning measurement outcome probabilities \cite{Be64,clauser69}, and hence could be tested in actual experiments, which systematically confirmed the validity of quantum predictions (see, e.g., Ref.\ \cite{christensen13} and references therein). Later, Bell's and EPR assumption of locality has been linked to the more general concept of ``non-contextuality''. The latter is the assumption that measurement results (even individual results, not only probabilities) for a given observable are independent of the choice of other compatible observables that are measured at (about) the same time \cite{compatible}. The importance of this generalization was highlighted by the Kochen-Specker (KS) theorem \cite{Ko67}, stating that for any physical system, prepared in any possible state (including separable states showing no special correlations, unlike EPR states), there exist a finite set of observables such that it is impossible to pre-assign them noncontextual results (i.e., independent of which other compatible observables are jointly measured) respecting the predictions of quantum theory.

The thought experiment ideated by Hardy in 1993 \cite{Ha93} -- ``one of the strangest and most beautiful gems yet to be found in the extraordinary soil of quantum mechanics'', in the words of David Mermin \cite{Me94} -- provides yet another way to highlight the conflict between the predictions of quantum mechanics and any non-contextual realistic interpretation. Like the Greenberger, Horne, and Zeilinger (GHZ) proposal \cite{GHZ90}, Hardy's paradox ideally relies only on testing a set of certainty or impossibility (``all versus nothing'') statements, as opposed to the quantitative assessment of probabilities needed for testing the original Bell's theorem, thus providing a form of ``Bell's theorem without inequalities''. On the other hand, Hardy's paradox has some advantages with respect to the GHZ one: only two, instead of three, compatible observables are required and only a partial, as opposed to maximal, degree of entanglement is needed. In fact, Hardy's paradox can be considered to be intermediate between the EPR-like paradoxes (including GHZ one), requiring a maximally entangled state, and the KS paradox, which applies to any state, including separable ones~\cite{Hardynote}.

A number of experimental tests of Hardy's paradox have been carried out hitherto \cite{torgerson95,digiuseppe97,boschi97,barbieri05,lundeen09,vallone11,fedrizzi11,chen12}, adopting different approaches, but all based on measuring properties of two separated particles (usually photons). Therefore, these tests mainly emphasized nonlocality, similarly to the experiments testing Bell-like inequalities. If we instead choose to focus on the more general concept of contextuality, we may also carry out experiments involving different degrees of freedom of a same particle. This approach has been taken already several times in the past, for experimentally testing the EPR, GHZ or KS paradoxes (see, e.g., \cite{Mi00,Hu03,Ha06,Ba09,Ka10,Damb13}), but to our knowledge it has not been applied before to Hardy's paradox. Here, we carry out a single-particle experimental test of Hardy's paradox by measuring two compatible degrees of freedom of the same photon, namely its spin angular momentum (SAM) and its orbital angular momentum (OAM). As for other experiments involving different degrees of freedom of a same particle, we cannot use locality to logically enforce non-contextuality, which must therefore be assumed, based only on the observables' compatibility (which might be tested with separate experiments). On the other hand, a single-photon experiment such as ours is simpler to implement and may allow much larger detection efficiencies than typical two-photon ones.

This paper is organized as follows. Section II recalls the basic idea of Hardy's paradox and illustrates its implementation in a spin-orbit four-dimensional Hilbert space of a single photon. Our experiment and its results are then described in Section III. Brief conclusions follow in Section IV. In addition, an Appendix is included with a proof of the statistical inequality that can be used to extend Hardy's paradox to the case in which experimental noise and imperfections preclude an all-versus-nothing test.

\section{Hardy's paradox for the polarization and OAM of a photon}
Hardy's paradox in our specific single-photon implementation can be sketched as follows. Alice and Bob must measure the properties of the photon and they agree on splitting the work: Alice will measure the photon polarization state and Bob its OAM. However, Alice can choose between two different polarization measurement setups, providing either one of two mutually incompatible observables that we will define in detail further on: for the time being, let us just call them $\Sigma$ and $\Sigma'$. In practice, Alice makes her setup choice by rotating a set of birefringent plates placed before a polarizing beam splitter (PBS). Similarly, Bob can choose between two different OAM measurement setups, providing either one of the two mutually incompatible observables $\Lambda$ and $\Lambda'$. Bob's choice is performed by selecting different hologram patterns visualized on a spatial light modulator (SLM). For both Alice and Bob, all observable measurements have only two possible outcomes, which will conventionally be labeled with the values $\pm1$. Alice's observable choice is supposed not to affect in any way Bob's measurement and vice versa, because the experimental procedures act on independent degrees of freedom of the photon. This corresponds to saying that $\Sigma$ and $\Sigma'$ are compatible with $\Lambda$ and $\Lambda'$.

Alice and Bob carry out a large number of measurements over an ensemble of identically prepared photons, choosing every time at random the observable to be measured for each degree of freedom. In this way, they estimate from the experimental frequencies the probabilities $P$ of different pairs of properties for the photon. In particular, they observe the following properties in their results:
\begin{enumerate}[label=P\arabic*]
\item \label{eq:P1} : The outcome $\Sigma=+1$ and $\Lambda=+1$ never occurs, that is $P_{\Sigma,\Lambda}(+1,+1)=0$
\item \label{eq:P2} :  The outcome $\Sigma=-1$ and $\Lambda'=-1$ never occurs, that is $P_{\Sigma,\Lambda'}(-1,-1)=0$
\item \label{eq:P3} : The outcome $\Sigma'=-1$ and $\Lambda=-1$ never occurs, that is $P_{\Sigma',\Lambda}(-1,-1)=0$
\end{enumerate}
As we shall prove further below, for any realistic non-contextual model, these three properties should logically imply the validity of the following fourth property:
\begin{enumerate}[label=P\arabic*,resume]
\item  \label{eq:P4} : The outcome $\Sigma'=-1$ and $\Lambda'=-1$ never occurs, that is $P_{\Sigma',\Lambda'}(-1,-1)=0$
\end{enumerate}
In contrast, Alice and Bob find that, in a sizable fraction of their measurements, the outcome $\Sigma'=-1$ and $\Lambda'=-1$ is indeed obtained, thus contradicting P4. This is Hardy's paradox.

\begin{figure}[ht]
	\centering
	\includegraphics[width=7cm]{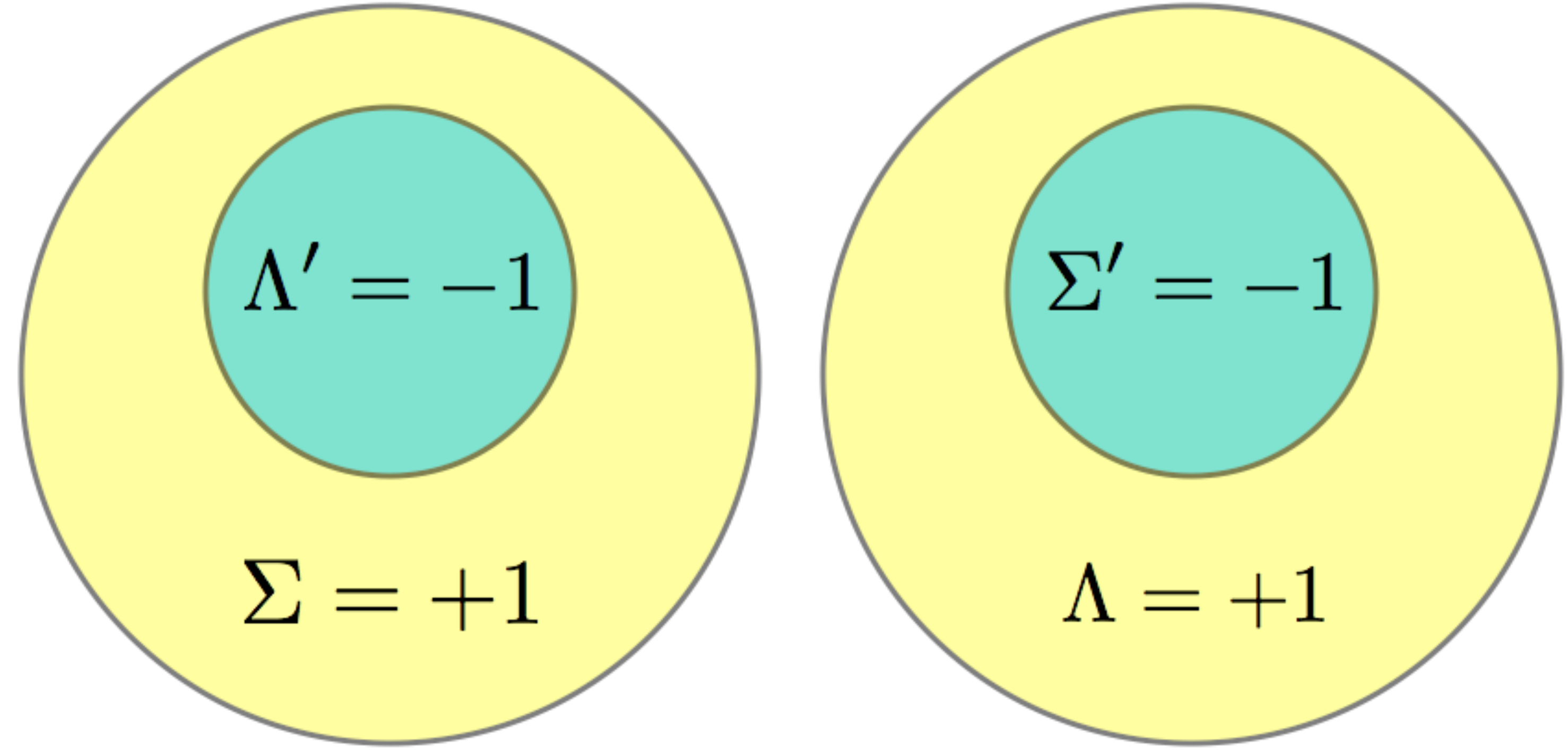}
	\caption{\label{fig:Venn}Venn diagrams for the events $\Sigma=+1, \Sigma'=-1, \Lambda=+1$, and $\Lambda'=-1$, satisfying conditions \ref{eq:P1}, \ref{eq:P2}, \ref{eq:P3}. The sets $\Sigma'=-1$ and $\Lambda'=-1$ being internal to disjoint sets cannot intersect, so that \ref{eq:P4} follows immediately.}
\end{figure}

The proof of \ref{eq:P4} from \ref{eq:P1}-\ref{eq:P3} is very simple. In a realistic non-contextual model, Alice's and Bob's measurement outcome events can be represented as Venn diagrams, which may be drawn on a plane that spans the underlying (hidden) reality determining all observable results. The Venn sets associated to the outcomes $\Sigma=+1, \Sigma'=-1, \Lambda=+1$, and $\Lambda'=-1$ are shown in Fig.~\ref{fig:Venn}. From property \ref{eq:P2} we infer that each time Bob measures $\Lambda'$, finding $\Lambda'=-1$, then if Alice measures $\Sigma$, she would certainly find $\Sigma=+1$. In other words, the event $\Lambda'=-1$ implies the event $\Sigma=+1$ which is represented by the fact that the Venn set for the event $\Lambda'=-1$ is internal to that for the event $\Sigma=-1$, as shown in Fig.~\ref{fig:Venn}. Similarly, from \ref{eq:P3} we infer that the event $\Sigma'=-1$ implies the event $\Lambda=+1$, that is, the Venn set of the event $\Sigma'=-1$ is internal to that of the event $\Lambda=+1$.
Now property \ref{eq:P1} implies that the Venn sets of the events $\Sigma=+1$ and $\Lambda=+1$ have no intersection, as shown in Fig.~\ref{fig:Venn}. It is now evident from the figure that the sets of the events $\Sigma'=-1$ and $\Lambda'=-1$ cannot intersect as well, from which we deduce property \ref{eq:P4}. It should be noted that the simultaneous presence of Venn sets for the results of incompatible observables, such as for example $\Sigma$ and $\Sigma'$, reveals a counterfactual aspect of the presented reasoning, because no single event can yield results for both observables, since they require different setups. However, we are actually assuming that, even when not measured, all observables values have a definite probability in each event.

\begin{figure*}[t]
	\centering
	\includegraphics[width=17cm]{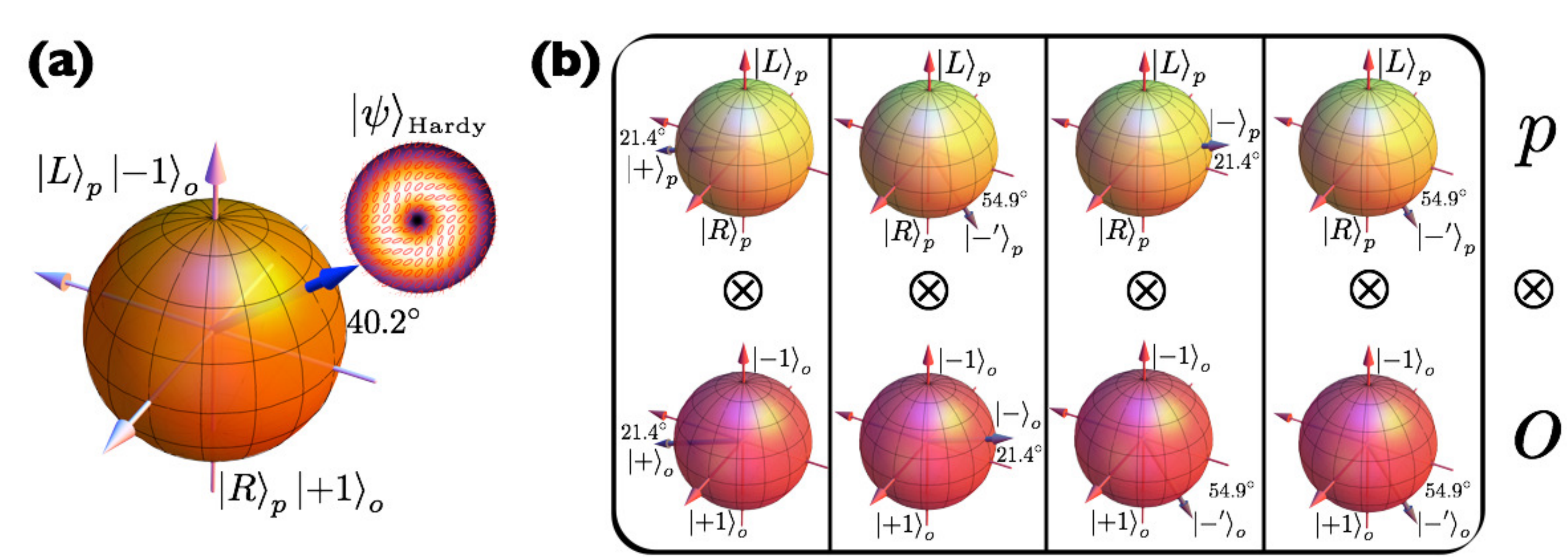}
	\caption{(a) Graphical representation of the photon Hardy state (\ref{eq:psi}) for $\gamma=24.9\ensuremath{^\circ}$ on the higher-order spin-orbit Poincar\'e sphere~\cite{milione11,holleczek:11}. The inset shows also the simulated transverse distribution of polarization (represented by the small ellipses) and intensity (represented by the color brightness) for the optical mode of such Hardy state. (b) Photon states in the Hilbert subspaces of polarization (upper row) and OAM (lower row), represented as arrows on Poincar\'e (Bloch) spheres. These states define the measured observables in Hardy's paradox test. The mathematical expressions of these states are given in Eq.(\ref{eq:plusminus}).} \label{fig:state}
\end{figure*}

Let us now illustrate the quantum mechanical description of the experiment and, based on that, we shall also specify the details of the state preparation and observable definitions to be used. Indeed, for a proper choice of states and observables, quantum mechanics can satisfy \ref{eq:P1}, \ref{eq:P2}, and \ref{eq:P3} and violate \ref{eq:P4}, in agreement with observations and in contradiction with a non-contextual realistic interpretation.

To start with, the photon must be prepared in the following partially entangled spin-orbit quantum state:
\begin{equation}\label{eq:psi}
   \ket{\psi}=\cos{(\gamma)}\ket{L}_p\ket{-1}_o-\sin{(\gamma)}\ket{R}_p\ket{+1}_o,
\end{equation}
where the $p$ and $o$ subscripts denote the polarization and OAM Hilbert spaces $\mathcal{H}_p$ and $\mathcal{H}_o$, respectively, $\ket{L}_p=\ket{+1}_p$ and $\ket{R}_p=\ket{-1}_p$ denote the left-handed and right-handed circular polarization states, corresponding also to the eigenstates of the SAM operator $\hat{S}_z$ with eigenvalues $\pm1$ (in units of $\hbar$), and $\ket{m}_o$ are the eigenstates of the OAM operator $\hat{L}_z$ with eigenvalue $m$ (also in units of $\hbar$), where $m$ is any integer. The angle $\gamma$ controls the degree of entanglement between the polarization and OAM degrees of freedom. The OAM Hilbert space is infinite-dimensional, but in this paper we shall restrict our attention to a two-dimensional subspace of $\mathcal{H}_o$ spanned by $m=\pm1$. In this way, $\mathcal{H}_p$ and $\mathcal{H}_o$ are isomorphic to each other and describe two standard qubit spaces, defined in separate degrees of freedom of the photon. We may think of state (\ref{eq:psi}) as a generic state in the four-dimensional spin-orbit Hilbert space $\mathcal{H}=\mathcal{H}_p\times\mathcal{H}_o$, up to ``local'' unitary operations acting separately on the polarization and OAM degrees of freedom. Only separable states (corresponding to $\gamma=n\pi/2$, with $n$ any integer) and maximally entangled states ($\gamma=\pi/4+n\pi/2$) cannot be used for implementing the Hardy paradox, as we shall see. In particular, in the following we shall restrict $\gamma$ to the range $0<\gamma<\pi/4$.

Let us now define the observables to be measured by Alice and Bob. In two-dimensional Hilbert spaces, all observables can be defined by giving a pair of orthogonal states that form a basis. So, let us first introduce the following four states in each qubit Hilbert subspace:
\begin{eqnarray}
	\label{eq:plusminus}
	\ket{+}_i &=& N \left(\sqrt{\sin{\gamma}}\;\ket{+1}_i+\sqrt{\cos{\gamma}}\;\ket{-1}_i\right) \nonumber \\
	\ket{-}_i &=& N \left(-\sqrt{\cos{\gamma}}\;\ket{+1}_i+\sqrt{\sin{\gamma}}\;\ket{-1}_i\right)\nonumber \\
	\ket{+'}_i &=& N' \left(\sqrt{\cos^3{\gamma}}\;\ket{+1}_i+\sqrt{\sin^3{\gamma}}\;\ket{-1}_i\right) \\
	\ket{-'}_i &=& N' \left(-\sqrt{\sin^{3}{\gamma}}\;\ket{+1}_i+\sqrt{\cos^{3}{\gamma}}\;\ket{-1}_i\right) \nonumber
\end{eqnarray}
where $N=(\sin\gamma+\cos\gamma)^{-1/2}$ and $N'=(\sin^3\gamma+\cos^3\gamma)^{-1/2}$, and $i=p,o$. The observable operators are then defined as follows:
\begin{eqnarray}
	\label{eq:observables}
	\hat{\Sigma}&=&\ket{+}_p \bra{+}_p - \ket{-}_p \bra{-}_p\nonumber \\
	\hat{\Sigma'}&=&\ket{+'}_p \bra{+'}_p - \ket{-'}_p \bra{-'}_p\nonumber \\
	\hat{\Lambda}&=&\ket{+}_o \bra{+}_o - \ket{-}_o \bra{-}_o \\
	\hat{\Lambda'}&=&\ket{+'}_o \bra{+'}_o - \ket{-'}_o \bra{-'}_o \nonumber
\end{eqnarray}
These definitions correspond to saying that in order to measure, for example, $\Sigma$, Alice must project the photon state onto $\ket{+}_p$ and $\ket{-}_p$. If the first projection (which is actually a filtering operation, in the experiment) succeeds, then the observable value is $+1$, otherwise it is $-1$. Similar procedures apply to all other observables.

Let us now calculate the quantum predictions for the four probabilities appearing in properties \ref{eq:P1}--\ref{eq:P4}. They are given by expressions such as $P_{\Sigma,\Lambda}(+1,+1)=|\bra{+}_p\bra{+}_o\ket{\psi}|^2$ and similar ones. A simple calculation shows that the probabilities appearing in \ref{eq:P1}, \ref{eq:P2}, \ref{eq:P3} are indeed zero, but that the probability in \ref{eq:P4} is given by
\begin{equation}\label{eq:Pnonzero}
   P_{\Sigma',\Lambda'}(-1,-1)=|\bra{-'}_p\bra{-'}_o\ket{\psi}|^2=\left[\frac{\sin 4\gamma}{4(\cos^3{\gamma}+\sin^3{\gamma})}\right]^2,
\end{equation}
which is nonzero for the range $0<\gamma<\pi/4$ (hence, for all partially entangled states, excluding only separable and maximally entangled ones). In particular, the probability (\ref{eq:Pnonzero}) is maximized for $\gamma\approx24.9\ensuremath{^\circ}$, for which $P_{\Sigma',\Lambda'}(-1,-1)=[(1+\sqrt{5})/2]^{-5}\approx9\%$. Hence, this is the best working point to test the paradox experimentally. The state given in Eq.\ (\ref{eq:psi}) with this choice for $\gamma$ can be represented by the vector drawn on the spin-orbit Poincar\'{e} sphere \cite{milione11,holleczek:11}, shown in Fig.~\ref{fig:state}a together with the transverse structure of the corresponding optical mode. The polarization and OAM states appearing in Eqs.\ (\ref{eq:plusminus}) are shown in Fig.\ \ref{fig:state}b, as vectors in the Poincar\'e (or Bloch) spheres representing the corresponding qubit Hilbert subspaces.

\section{Experiment}
\begin{figure}[ht]
	\centering
	\includegraphics[width=9cm]{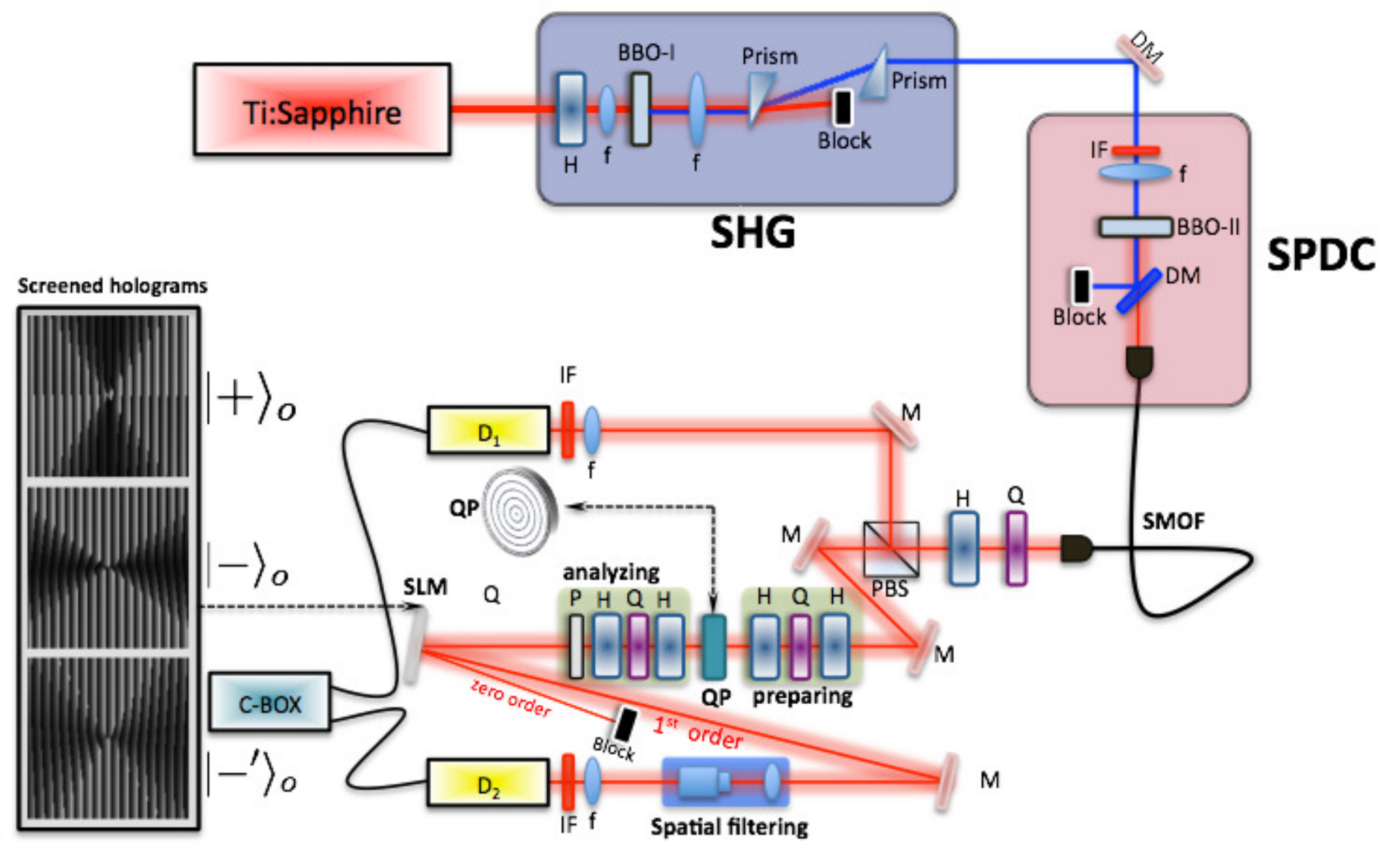}
	\caption{Experimental setup. The SHG and SPDC stages are used to prepare the heralded photon used in the Hardy test and its trigger companion, as explained in the main text. In the main setup, the heralded photon is prepared in a controlled polarization state and then converted into a spin-orbit partially entangled state by the q-plate. For carrying out the tomography of this state and the Hardy test, the photon was then projected (filtered) with another sequence of wave plates and a polarizer, for the polarization measurement (Alice), and by diffraction on a computer-generated holograms (hologram patterns used for filtering specific states are shown in the left inset) screened on a SLM, followed by spatial filtering in zero OAM mode and photon detection. The photon counts coming from the trigger detector (D$_1$) and the heralded photon one (D$_2$) were sent to a coincidence box, for recording the number of photon coincidences occurring in a given time window. Legend: C-BOX - coincidence box; D$_i$ - single photon detector; DM - dichroic mirror; f - lens; H - half-wave plate; IF - bandpass interference filter; M - mirror; Q - quarter-wave plate; QP - q-plate; P - polarizer; SLM - spatial light modulator.}
	\label{fig:setup}
\end{figure}

The full layout of the experimental apparatus of our experiment is shown in Fig.\ \ref{fig:setup}. The pump pulses are emitted at 82 MHz repetition rate by a Ti:sapphire mode-locked laser, centered at the wavelength $\lambda=795$ nm. The pulses are frequency-doubled by second-harmonic generation (SHG) in a first $\beta$-barium borate (BBO) nonlinear crystal, cut for type I phase matching, and then used to generate photon pairs by spontaneous parametric down conversion (SPDC) in a second BBO nonlinear crystal, cut for type II degenerate phase matching. The photons were coupled to a single-mode optical fiber to clean the spatial mode and sent to the main apparatus. After compensating for the polarization rotations induced by the fiber, the photons are split by a PBS, and one photon is detected by a silicon avalanche photodiode (AD) and used as trigger. The other photon is used for the Hardy's test, in a heralded single-photon regime.

In order to prepare the photon input state, we used the q-plate technology~\cite{Ma06,Slu11}, which provides a simple and convenient method for entangling polarization and OAM degrees of freedom \cite{Nag09prl}. In this work, for the first time we exploit a q-plate with $q=1/2$ to generate partially entangled states, rather than maximally entangled ones. A q-plate, tuned electrically to maximal conversion efficiency \cite{Pic10}, operates the spin-orbit optical transformations $\ket{L}_p\ket{0}_o\rightarrow\ket{R}_p\ket{+1}_o$ and $\ket{R}_p\ket{0}_o\rightarrow\ket{L}_p\ket{-1}_o$. Hence, in order to generate the state (\ref{eq:psi}), we need only to prepare the input photon in the zero OAM mode, as obtained after passing through a single-mode fiber, and with elliptical polarization $\cos{(\gamma)}\ket{R}_p-\sin{(\gamma)}\ket{L}_p$ and then pass it through the q-plate. The parameter $\gamma$, fixing the degree of entanglement in the final state, was adjusted using a sequence of birefringent wave plates: a half-wave plate, with optical axis rotated at the angle $\gamma/2$ from the direction of the input polarization, a quarter-wave plate, with the axis set at $\pi/4$, and another half-wave plate at $-\pi/8$ (see Fig. \ref{fig:setup}).

\begin{figure}[ht]
	\centering
	\includegraphics[width=8cm]{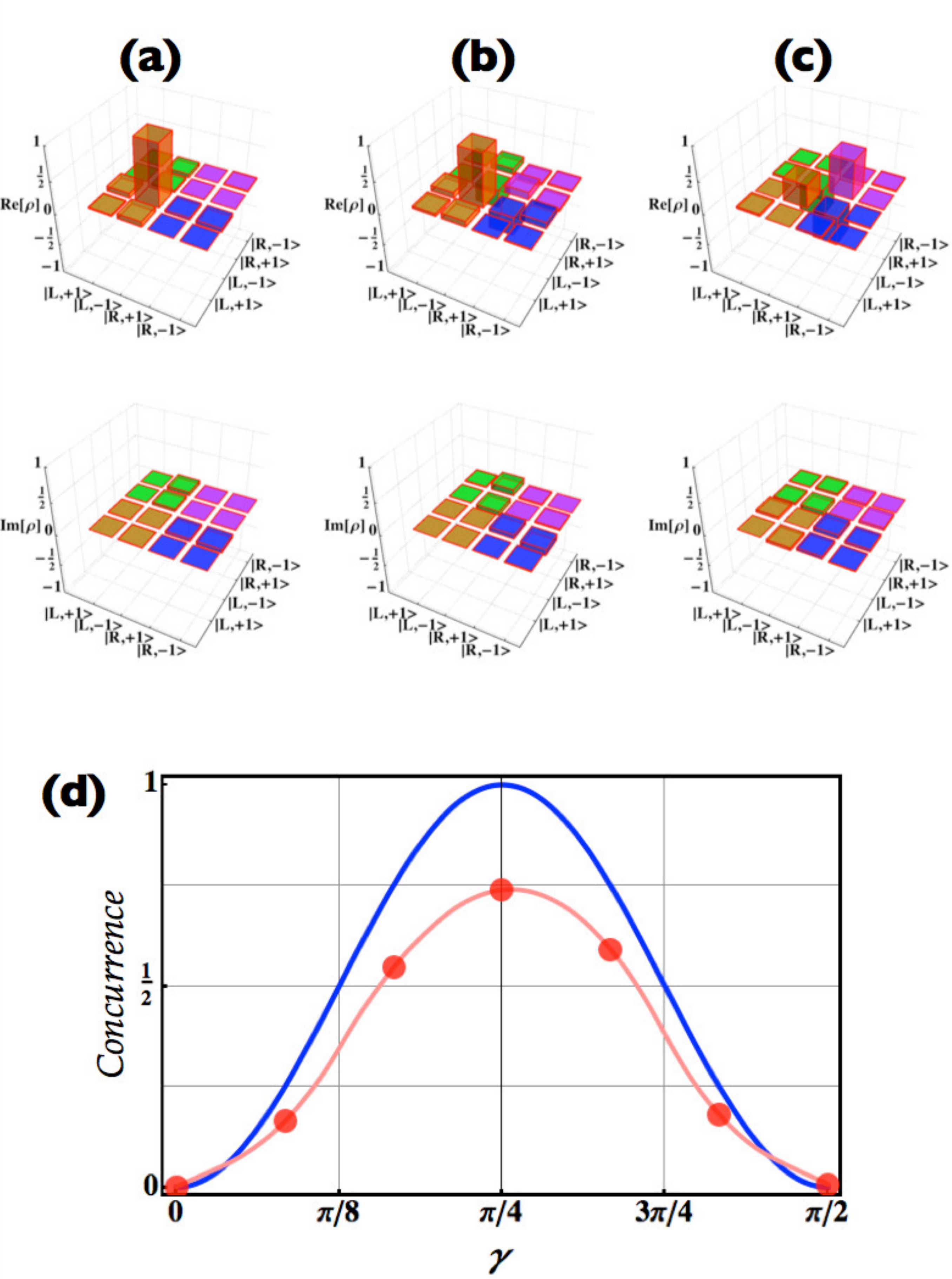}
	\caption{Experimental characterization of partially entangled spin-orbit single-photon states defined in Eq.\ (\ref{eq:psi}). (a-c) Experimental reconstructed density matrix for three different values of the entanglement parameter $\gamma$: (a) $\gamma=0$, (b) $\gamma=\pi/8$, (c) $\gamma=\pi/4$. The associated state fidelities are $F=0.985\pm0.004$, $F=0.984\pm0.004$, and $F=0.956\pm0.005$, respectively. (d) Concurrence value of Schmidt's state as a function of $\gamma$. The red points are the experimental data and the red solid line is an interpolating function used as a guide for the eye, while the solid blue line is from theory. The concurrence value for the Hardy state at the optimal angle $\gamma=24.9\ensuremath{^\circ}$, is equal to 0.36.}
	\label{fig:tomography}
\end{figure}

We checked the quality of the prepared state by carrying out a full quantum tomography, based on projecting (filtering) the state on a set of mutually unbiased bases in both polarization and OAM Hilbert spaces and then using a standard maximal likelihood estimator for best fit. The polarization filtering was carried out by using a second set of birefringent wave plates followed by a polarizer. The OAM filtering was based on a standard holographic method \cite{Nag09opex}, by diffracting the photons on a set of computer-generated holograms (optimized for fidelity \cite{Bold13,Damb13sr}) visualized on a SLM and then filtering the zero OAM component within the first-order diffracted photons. Fig.\ \ref{fig:tomography}(a-c) shows the experimentally reconstructed density matrix $\op{\rho}_\psi=\ket{\psi}\bra{\psi}$ of the generated spin-orbit state (\ref{eq:psi}) for $\gamma=0$, $\gamma=\pi/8$, and $\gamma=\pi/4$. The corresponding concurrence, as a measure of the degree of entanglement \cite{concurrence}, was also determined for seven values of $\gamma$ and the results are shown in Fig.\  \ref{fig:tomography}(d).

\begin{figure}[th]
	\centering
	\includegraphics[width=8cm]{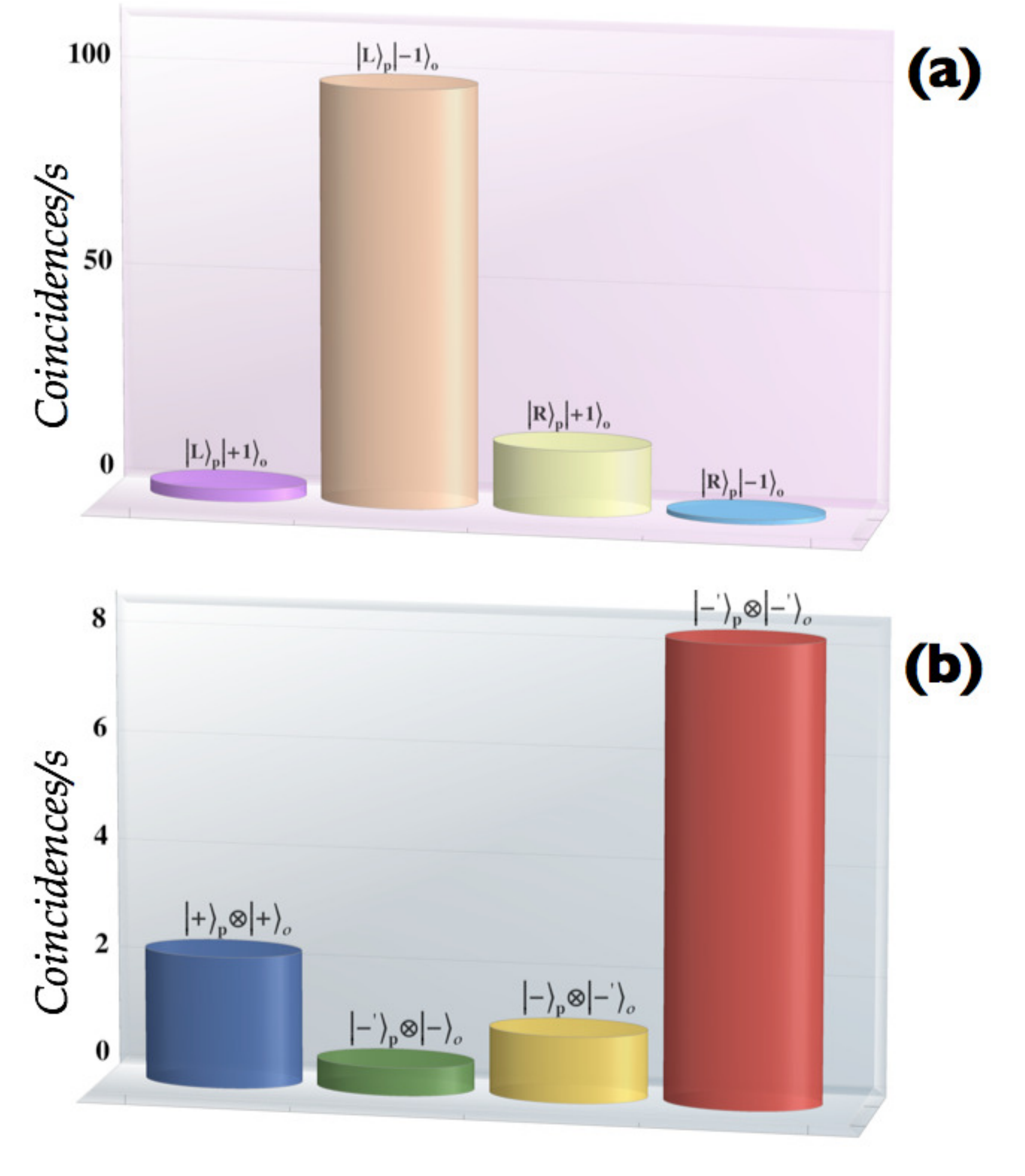}
	\caption{Experimental verification of Hardy's paradox, by using state (\ref{eq:psi}) for the input photon with $\gamma=24.9\ensuremath{^\circ}$. (a) Coincidence count rates for the four spin-orbit basis states. The measurement time window was of 100 s and the mean total number of coincidences was of $N_{tot}=12000\pm110$, corresponding to the rate $C_{tot}=120.0\pm1.1$ coincidences/s. (b) Coincidence count rates for the measurement of spin-orbit states (\ref{eq:plusminus}), which define the observables $\Sigma, \Sigma', \Lambda$, and $\Lambda'$. The ratio of the reported coincidence rate values with $C_{tot}$ gives the experimental frequencies and the estimates of the probabilities for the photon to be in the corresponding state.}
\label{fig:paradox}
\end{figure}

For carrying out the Hardy test, we then set $\gamma=24.9\ensuremath{^\circ}$ in order to maximize the probability (\ref{eq:Pnonzero}), as explained above. While heralded photons prepared in this state are sent in the main apparatus, we made a series of projective measurements of the observables $S_z, \Sigma, \Sigma'$ in the polarization subspace and $L_z, \Lambda,\Lambda'$ in the OAM subspace. In particular, the probabilities of detecting specified spin-orbit states were assessed experimentally by counting the number of photon coincidences between the test photon and the trigger one in a given time window of 100 s. We first measured the probabilities of the four spin-orbit basis states $\ket{L}_p\ket{+1}_o, \ket{L}_p\ket{-1}_o, \ket{R}_p\ket{+1}_o, \ket{R}_p\ket{-1}_o$, which are eigenstates of $\hat{S}_z$ and $\hat{L}_z$. The results are shown in Fig.\ \ref{fig:paradox}a and are in reasonable agreement with the quantum predictions obtained from state (\ref{eq:psi}). More precisely, while the counts for states $\ket{L}_p\ket{-1}_o$ and $\ket{R}_p\ket{-1}_o$ were consistent with theory within experimental uncertainties, the counts for state $\ket{R}_p\ket{+1}_o$  were about 25\% smaller than theory and there was a 3\% fraction of counts for state $\ket{L}_p\ket{+1}_o$, which in theory should have vanishing probability.

Next, we measured the four probabilities appearing in properties \ref{eq:P1}-\ref{eq:P4} entering Hardy's paradox, by performing a projective measurement on the four states $\ket{+}_p\ket{+}_o, \ket{-'}_p\ket{-}_o, \ket{-}_p\ket{-'}_o$, and $\ket{-'}_p\ket{-'}_o$. The experimental results are given in Fig.\ \ref{fig:paradox}b. The count frequency of state $\ket{-'}_p\ket{-'}_o$ was found to be $(7.4\pm0.2)$\% (specified errors are estimated as standard deviations computed assuming Poissonian statistics and ignoring other possible sources of errors), against a quantum prediction of 9\%. The other states presented much smaller, but nonvanishing count frequencies: $(2.1\pm0.1)$\%, $(0.45\pm0.06)$\%, and $(1.0\pm0.1)$\%, respectively for states $\ket{+}_p\ket{+}_o, \ket{-'}_p\ket{-}_o$, and  $\ket{-}_p\ket{-'}_o$. This outcome, probably due to an imperfect state preparation and/or to some residual cross-talk in the OAM and polarization measurements, makes it not possible to apply the simple all-versus-nothing reasoning presented above for Hardy's paradox. But this is normal for an experimental test, as no experimental result can be perfectly zero, because of the unavoidable noise and other experimental imperfections.

To take into account experimental imperfections, one must replace the all-versus-nothing paradox with a Bell-like inequality, as for example discussed by Mermin~\cite{Me94} (on this issue, see also \cite{Ghi08}). More precisely, the inequality arising in Hardy's paradox is the following one (see Appendix A for a proof):
\begin{eqnarray}\label{eq:inequality}
P_{\Sigma',\Lambda'}(-1,-1) &\le& P_{\Sigma,\Lambda}(+1,+1)+P_{\Sigma',\Lambda}(-1,-1)\nonumber\\
&&+P_{\Sigma,\Lambda'}(-1,-1)  
\end{eqnarray}
This inequality should hold true in any non-contextual realistic model and is trivially satisfied in the ideal noiseless case given by properties \ref{eq:P1}--\ref{eq:P4}. But it is violated by quantum predictions. Our results reported above also violate the inequality by over seven standard deviations, thus confirming the contextual behavior of quantum mechanics.

\section{Conclusions}
In synthesis, we have demonstrated generation of partially entangled spin-orbit states of single photons and have exploited such states to carry out an experimental test of Hardy's paradox. Although ideally Hardy's paradox is an all-versus-nothing statement, so that a single event could be enough to prove that its assumptions must be wrong, in practice experimental noise forces the use of an inequality and of a statistical verification, similar to the case of a Bell inequality test. In our experiment, we verified that the inequality was violated with high statistical confidence. This proves that a non-contextual realistic model of the system cannot be true. Non-contextuality in our single particle experiment was assumed based on the observable compatibility, and not on their spatial separation. While this is a conceptual limitation of our test, the use of a single particle demonstrated in our approach has some other advantages. The first is that the experiment is much simpler to implement. We used a heralded photon regime, but a similar demonstration could be based on an attenuated light source, thus making it even simpler. Simple-to-perform experiments testing the conceptual paradoxes of quantum mechanics may have a strong educational value. A second advantage is that the quantum detection efficiency for a single-photon detection can be made significantly larger than for the two-photon case. This, in turn, could be exploited in the future to carry out Hardy-like tests that are free from other assumptions, such as the fair sampling hypothesis, or other detection-related loopholes that may undermine their validity, similarly to the case of Bell inequalities (on this issue see, e.g., Refs.\ \cite{christensen13,Sant04,Gerh11}).

\section*{Acknowledgements}
This work was supported by the 7th Framework Programme of the European Commission, within the Future Emerging Technologies program, under Grant No. 255914, PHORBITECH.

\appendix
\section{Proof of Hardy inequality}
We derive here the inequality (\ref{eq:inequality}), which is to be satisfied by any realistic and non-contextual statistical model for Hardy's paradox. Realism of the model means that we can assign probabilities to the set of elementary events corresponding to each possible result of the measurement of the observables $\Sigma,\Lambda,\Sigma',\Lambda'$, even when they are not measured. Because each measurement can have only the two results $\pm 1$, we have sixteen probabilities associated to each event. The difference between a realistic statistical model and quantum mechanics stems from the fact that, in the former, probabilities can be assigned to any $\sigma$-algebra of the elementary events, while this is impossible in quantum mechanics in general. For example, any realistic statistical model assigns values (eventually zero or one) to all 16 probabilities ${\cal P}_n$ $(n=1,\dots,16)$ of the four-fold joint measurements of $\Sigma$, $\Sigma'$, $\Lambda$, and $\Lambda'$, as reported in Table~\ref{tab:1}, which is impossible in quantum mechanics, because the primed observables do not commute with the unprimed ones.
\begin{table}[ht]
   \begin{tabular}{|l |c |c |c |c ||l |c |c |c |c |}
      \hline
                      & $\Sigma$  & $\Sigma'$  & $\Lambda$  & $\Lambda'$   &                     & $\Sigma$  & $\Sigma'$  & $\Lambda$  & $\Lambda'$  \\
      \hline
      ${\cal P}_1$    &  $-1$  &  $-1$   &     $-1$   &    $-1$      &     ${\cal P}_9$    &  $+1$  &  $-1$   &     $-1$   &    $-1$     \\
      ${\cal P}_2$    &  $-1$  &  $-1$   &     $-1$   &    $+1$      &     ${\cal P}_{10}$ &  $+1$  &  $-1$   &     $-1$   &    $+1$     \\
      ${\cal P}_3$    &  $-1$  &  $-1$   &     $+1$   &    $-1$      &     ${\cal P}_{11}$ &  $+1$  &  $-1$   &     $+1$   &    $-1$     \\
      ${\cal P}_4$    &  $-1$  &  $-1$   &     $+1$   &    $+1$      &     ${\cal P}_{12}$ &  $+1$  &  $-1$   &     $+1$   &    $+1$     \\
      ${\cal P}_5$    &  $-1$  &  $+1$   &     $-1$   &    $-1$      &     ${\cal P}_{13}$ &  $+1$  &  $+1$   &     $-1$   &    $-1$     \\
      ${\cal P}_6$    &  $-1$  &  $+1$   &     $-1$   &    $+1$      &     ${\cal P}_{14}$ &  $+1$  &  $+1$   &     $-1$   &    $+1$     \\
      ${\cal P}_7$    &  $-1$  &  $+1$   &     $+1$   &    $-1$      &     ${\cal P}_{15}$ &  $+1$  &  $+1$   &     $+1$   &    $-1$     \\
      ${\cal P}_8$    &  $-1$  &  $+1$   &     $+1$   &    $+1$      &     ${\cal P}_{16}$ &  $+1$  &  $+1$   &     $+1$   &    $+1$     \\
      \hline
   \end{tabular}
   \caption{Set of all possible values for the four observables $\Sigma$, $\Sigma'$, $\Lambda$, $\Lambda'$ and corresponding symbols for the probabilities.}\label{tab:1}
\end{table}
Besides realism, the model non-contextuality corresponds to the assumption that the results of a measurements on a given observable are independent of possible joint measurements carried out on other compatible observables. This means that each probability ${\cal P}_n$ depends on the observables pertaining the system under study and not on the context in which the measurements are performed. In particular, the ${\cal P}_n$ cannot depend on parameters characterizing the measurement apparatus or the environment. Non-contextuality is always assumed in classical physics and it is implicit in the probabilities ${\cal P}_n$ in Table~\ref{tab:1}, independently of their actual values. From the probabilities ${\cal P}_n$, we may easily calculate the probabilities defining properties \ref{eq:P1}--\ref{eq:P4}, relevant for Hardy's paradox. We obtain
\begin{eqnarray}\label{eq:P1P4}
   P_{\Sigma,\Lambda}(+1,+1) &=& {\cal P}_{11} + {\cal P}_{12} +{\cal P}_{15} +{\cal P}_{16}  \nonumber \\
   P_{\Sigma',\Lambda}(-1,-1) &=& {\cal P}_{1} + {\cal P}_{2} +{\cal P}_{9} +{\cal P}_{10}  \nonumber \\
   P_{\Sigma,\Lambda'}(-1,-1) &=& {\cal P}_{1} + {\cal P}_{3} +{\cal P}_{5} +{\cal P}_{7} \nonumber \\
   P_{\Sigma',\Lambda'}(-1,-1) &=& {\cal P}_{1} + {\cal P}_{3} +{\cal P}_{9} +{\cal P}_{11}
\end{eqnarray}
Observing that all probabilities on the right of the last of Eqs.~(\ref{eq:P1P4}) are already present on the right of the first three equations, we find inequality (\ref{eq:inequality}).
%
%


\begin{thebibliography}{99}
\bibitem{compatible}
The compatibility of two observables $A$ and $B$ can be operationally defined by the observed independence of their measurement results from the order by which we measure them and the possible measurement repetition. For example, if $A$ and $B$ are two compatible observables, we can perform the measurement sequences $AB$, $BA$, $ABA$, $BAB$, etc.\ and we would always obtain the same results for the values of $A$ and $B$. In the quantum formalism, observables are compatible if and only if their associated operators commute.
\bibitem{Hardynote}
However, unlike KS paradox, Hardy's paradox requires measurements on observables which are state-dependent.
\bibitem{concurrence}
The concurrence of a generic state $\rho_\psi$ of a bipartite quantum system is defined as $C(\rho_\psi)=\min\{0,\lambda_1-\lambda_2-\lambda_3-\lambda_4\}$, where$\lambda_i$ are the square roots of the eigenvalues of $\bar{\rho}_\psi\cdot\rho_\psi$ where $\bar{\rho}_\psi=(\sigma_y \otimes\sigma_y)\cdot{\rho_\psi}^{\star}\cdot(\sigma_y \otimes\sigma_y)$ with $\sigma_y$ Pauli matrix.
\end{thebibliography}
\end{document}